\documentclass[twocolumn,aps,prl,showpacs,floatfix]{revtex4}
\usepackage{color} %CB to be deleted
\usepackage[normalem]{ulem}  %CB to be deleted

\usepackage{graphicx}
\usepackage{times}
\usepackage{nicefrac}
\usepackage{amsmath}
\usepackage{amsfonts}
\usepackage{amssymb}
\usepackage{amsthm}
\usepackage{epsf}
\usepackage{bm}
\usepackage{times}

\usepackage{dcolumn}

\newcolumntype{.}{D{x}{}{-1}}

\newcommand{\bfB}{{\bf B}}

\newcommand{\be}{\begin{eqnarray}}
\newcommand{\ee}{\end{eqnarray}}

\newcommand{\mub}{\mu_\text{B}}
\newcommand{\mun}{\mu_\text{N}}
\newcommand{\DEnuc}{\Delta E_\text{nuc}}
\newcommand{\DEmagn}{\Delta E_\text{magn}}
\newcommand{\DEhfs}{\Delta E_\text{HFS}}
\newcommand{\dghfs}{\delta g_\text{HFS}}

\newcommand{\BWu}{B_\text{W.u.}}
\newcommand{\gfe}{g_\text{e}}
\newcommand{\gfexp}{g^\text{exp}}
\newcommand{\gfeth}{\gfe^\text{th}}
\newcommand{\masse}{m_\text{e}}
\newcommand{\massp}{m_\text{p}}

\newcommand{\bmu}{{\mbox{\boldmath$\mu$}}}
\newcommand{\mf}{M_F}

\begin{document}
%%%%%%%%%%%%%%%%%%%%%%%%%%%%%%%%%%%%%%%%%%%%%%%%%%%%%%%%%%%%%%%%%%%%

\title{Ground-state $g$ factor of highly charged $^{229}$Th ions:\\
  an access to the M1 transition probability between the isomeric and ground
  nuclear states}

\author{V.~M.~Shabaev$^1$, D.~A.~Glazov$^1$, A.~M.~Ryzhkov$^1$, C.~Brandau$^{2,3}$, G.~Plunien$^4$,
W.~Quint$^3$, A.~M.~Volchkova$^1$, and D.~V.~Zinenko$^1$}

\affiliation{
$^1$Department of Physics, St.Petersburg State University, Universitetskaya 7/9, 199034 St.Petersburg, Russia\\
$^2$I. Physikalisches Institut, Justus-Liebig-Universit\"at, Heinrich-Buff-Ring 16, D-35392 Giessen, Germany\\
$^3$GSI Helmholtzzentrum f\"ur Schwerionenforschung GmbH, Planckstrasse 1, D-64291 Darmstadt, Germany\\
$^4$Institut f\"ur Theoretische Physik, TU Dresden, Mommsenstrasse 13, Dresden, D-01062, Germany}

\begin{abstract}
A method is proposed to determine the $M1$ nuclear transition amplitude and hence the lifetime 
of the ``nuclear clock transition'' between the low-lying  ($\sim 8$ eV) first isomeric state 
and the ground state of $^{229}$Th from a measurement of the \textit{ground-state} $g$ factor 
of few-electron $^{229}$Th ions. As a tool, the effect of nuclear hyperfine mixing (NHM)
in highly charged $^{229}$Th-ions such as $^{229}$Th$^{89+}$ or $^{229}$Th$^{87+}$ is utilized.
The ground-state-only $g$-factor measurement would also provide first experimental evidence of NHM in atomic ions.
Combining the measurements for H-, Li-, and B-like  $^{229}$Th ions has a potential to improve 
the initial result for a single charge state and to determine the nuclear magnetic moment 
to a higher accuracy than that of the currently accepted value. 
The calculations include relativistic, interelectronic-interaction, QED, and nuclear effects.

\end{abstract}
%\pacs{31.30.J-, 12.20.Ds}
\maketitle

%%%%%%%%%%%%%%%%%%%%%%%%%%%%%%%%%%%%%%%%%%%%%%%%%%%%%%%%%%%%%%%%%%%%
% \section{Introduction}
%%%%%%%%%%%%%%%%%%%%%%%%%%%%%%%%%%%%%%%%%%%%%%%%%%%%%%%%%%%%%%%%%%%%%

% \textit{Introduction.} ---
The exceptionally low-energy (about 8 eV) isomeric state in $^{229}$Th,
which is connected to the ground state by a magnetic dipole (M1) transition, 
attracts great interest of metrology institutes as well as atomic and nuclear physics communities
\cite{bec07,sei19,mei19,yam19,bor19,sik20,pei21,bee21} worldwide. 
Among others, $^{229}$Th is considered as an ideal testbed of temporal variations of fundamental constants, 
as a nuclear $\gamma$-ray laser \cite{tka11} or as an ideal candidate for a nuclear-transition based optical clock 
that eventually could serve as a new metrological frequency standard with unrivaled properties \cite{pei03,cam12,pei15}.
The practical realization of these applications requires the precise knowledge of the excitation energy 
as well as other fundamental nuclear properties such as nuclear magnetic moments of the ground state (g.s.) 
and the isomeric state (i.s.), and, as a key property of a clock, the lifetime of the isomer.
The excitation energy was measured to an accuracy of about 2\%
(8.28(17) eV in Ref. \cite{sei19} and 8.10(17) eV in Ref. \cite{sik20})
and the magnetic moments of g.s. and i.s. were derived
from experiments to precision of about 2\%  and 16\%, respectively \cite{cam11,saf13,thi18}.
The lifetime of the neutral $^{229}$Th is dominated by internal conversion
which is more than 9 orders of magnitude stronger than the gamma decay \cite{kar07,tka15}.
Up to now, the internal conversion is also the only approach for direct detection of the isomer
transition since neither the direct excitation nor the decay photons could be observed yet.
Thus, to date there is no accurate experimental data on the $M1$ transition probability between these states
(in accordance with Ref. \cite{tka15}, we neglect higher-order multipole contributions to the isomeric decay rate).
For $^{229}$Th$^{2+}$ in which internal conversion is energetically not possible, an experimental lower bound for
the lifetime of the gamma-decay of 1 min was reported \cite{wen16}.
Calculations of the reduced transition probability $B(M1)$ span the range from 0.005 to 0.048 Weisskopf units (W.u.)
\cite{dyk98,ruc06,tka15,min17,min19,min21}. A very recent indirect estimation of $B(M1)$ from half-life measurements 
of other nuclear excited states in $^{229}$Th yields 0.008(2) W.u.~\cite{shigekawa:21:prc}. This estimation agrees
with the most elaborated theoretical predictions, $0.006$--$0.008$ W.u., of Refs.~\cite{min19,min21}.
Yet, a precise experimental determination of this important value is still pending.

In the present Letter,
we propose a method for a highly sensitive experimental determination of the $^{229}$Th transition probability that
is deduced from a measurement of the $g$ factor of highly charged $^{229}$Th$^{q+}$ in its g.s.
In few-electron $^{229}$Th the most tightly bound unpaired electron produces a strong magnetic field 
at the site of the nucleus and leads to a nuclear hyperfine mixing (NHM) of the states.
The mixing coefficient $b$ enters the $g$ factor of the ion and contains the information 
of the $M1$-transition probability. Hence, the decay property of the i.s. can be experimentally deduced 
from an ion that is in the nuclear g.s. To date, measurements of the $g$ factor of H- and Li-like ions 
with the nuclear charge number $Z=6$--$20$ \cite{haf00,ver04,stu11,wag13,lin13,stu13,stu14,koel16,gla19}
have reached an accuracy of about $3\times 10^{-10}$ or better. It is expected that the same accuracy 
will be achieved in $g$-factor experiments with very heavy few-electron ions at the highly charged ion trap 
facility HITRAP at the accelerator complex of GSI/FAIR in Darmstadt, Germany \cite{klu08,her15}.
Alternatively, such high charge states can be produced at electron beam ion traps \cite{bei93,bei98,gon05}.
We show that the experimental determination of the ground-state $g$ factor of H-like  $^{229}$Th ion to the precision
of about $10^{-7}$ allows one to get the NHM mixing coefficient $b$ to an accuracy of about $10^{-3}$.
Using this value of $b$ and the excitation energy $\DEnuc$ known from the experiments \cite{sei19,sik20}, 
one can get $B(M1)$ with a few-percent accuracy.
Furthermore, a comparison of the measurements of the ground-state $g$ factors
of H-, Li-like and B-like $^{229}$Th ions improves the $b$ value by about one order
of magnitude and allows precise determination of the nuclear magnetic moment.

The approach is based on the NHM effect in highly charged $^{229}$Th ions \cite{lyu66,sze90,wyc93,kar98,pac01,bel14,tka16a,tka16b}. 
NHM is most pronounced in one-electron $^{229}$Th$^{89+}$, three-electron $^{229}$Th$^{87+}$, 
or five-electron $^{229}$Th$^{85+}$ with an unpaired valence $j= 1/2$ electron.
In these charge states, in addition to the ordinary hyperfine structure, 
the very strong magnetic field of up to $\sim$ 28~MT ($^{229}$Th$^{89+}$) of the unpaired electron 
mediates a mixing of the $F = 2$ levels of the g.s. and i.s., 
i.e. a mixing of nuclear ground and isomeric levels with the same electronic state. 
In contrast to NHM, hyperfine mixing of electronic states (often termed hyperfine quenching)
has been studied theoretically as well as experimentally for a large number of atomic metastable ions 
with charge states ranging from neutral atoms or singly charged ions \cite{bec01,boy07,ros07}
up to extreme cases such as two-electron $^{155}$Gd$^{62+}$ and $^{157}$Gd$^{62+}$ \cite{ind92} or $^{197}$Au$^{77+}$ \cite{tol04}.

NHM results in an additional small energy shift, 
but more notable, the lifetime of the i.s. decreases drastically, for $^{229}$Th$^{89+}$
by up to 5-6 orders of magnitude, from a few hours down to a few tens of ms.
It is noted that due to this vast increase of the transition rates NHM might also become a key asset
for laser spectroscopy of the nuclear transition. In fact, the experimental parameters become
similar to the ones of successful storage-ring laser experiments of ordinary hyperfine transitions
in $^{209}$Bi$^{82+}$ and $^{209}$Bi$^{80+}$ \cite{kla94,ull17}. 

The mixing coefficient $b$ is a function of the nuclear excitation energy $\DEnuc$ and the transition probability  $B(M1)$.
In the case of small mixing, it can be approximated as $b\sim \sqrt{B(M1)}/\Delta E_{\rm nuc}$,
where the proportionality coefficient can be calculated to a good accuracy for a given ion.
NHM is well known for muonic atoms (see, e.g., Refs. \cite{wu69,hit70,mic19} and references therein), but has not been measured in conventional atoms or ions by now.
Thus, the proposed $g$-factor measurements of few-electron $^{229}$Th would also provide experimental evidence of the electronic NHM effect.

For a $^{229}$Th$^{q+}$ g.s. ion ($I^{\pi}=5/2^+$) with a single $j=1/2$ valence electron
the hyperfine interaction splits the g.s. of the ion into two sublevels with the total angular momentum
$F=2$ and $F=3$. Similarly, the i.s. ($I^{\pi}=3/2^+$) splits into sublevels with $F=1$ and $F=2$.
Due to the NHM, the $F=2$ states can be represented as
\begin{align}
  \overline{|5/2^+,F=2\rangle} =& \sqrt{1-b^2}\, |5/2^+,F=2\rangle
\nonumber\\
    & - b\, |3/2^+,F=2\rangle
\label{state_ground}
\,,\\
  \overline{|3/2^+,F=2\rangle} =& \sqrt{1-b^2}\, |3/2^+,F=2\rangle
\nonumber\\
    & + b\, |5/2^+,F=2\rangle
\,.
\label{state_isomeric}
\end{align}
The NHM coefficient $b$ can be determined from
\begin{equation}
\label{b}
  b^2= \frac{1}{2} - \frac{1}{2}\frac{|E_1-E_2|}{\sqrt{(E_1-E_2)^2+4|V_{21}|^2}}
\,,
\end{equation}
where $E_1=E_1^0+V_{11}$ and $E_2=E_2^0+V_{22}$ are the energies
of the $F=2$ g.s. and i.s. ions neglecting the mixing effect,
$E_{1,2}^0$ are the energies in the absence of the hyperfine splitting 
(we choose $E_1^0=0$ and hence $E_2^0=\DEnuc$),
$V_{11}$ and $V_{22}$ are the corresponding expectation values of the hyperfine interaction,
and $V_{21}$ is the nondiagonal matrix element of the hyperfine interaction.
The energies including the NHM effect are given by
\begin{equation}
\label{e12}
  \overline{E}_{1,2}=\frac{E_1+E_2}{2} \mp \frac{1}{2} {\sqrt{(E_1-E_2)^2+4|V_{21}|^2}}\,.
\end{equation}
In the case of small mixing ($b \ll 1$), expanding Eqs. (\ref{b}) and (\ref{e12}) 
in the parameter $V_{21}/(E_1-E_2)$, we find
\begin{align}
\label{b_ap}
  b & \approx b_0 \equiv -\frac{V_{21}}{E_1-E_2}
\,,\\
  \overline{E}_{1,2} & \approx E_{1,2} \pm \frac{|V_{21}|^2}{E_1-E_2}
\,.
\end{align}
\begin{figure}
\begin{center}
  \includegraphics[width=\columnwidth]{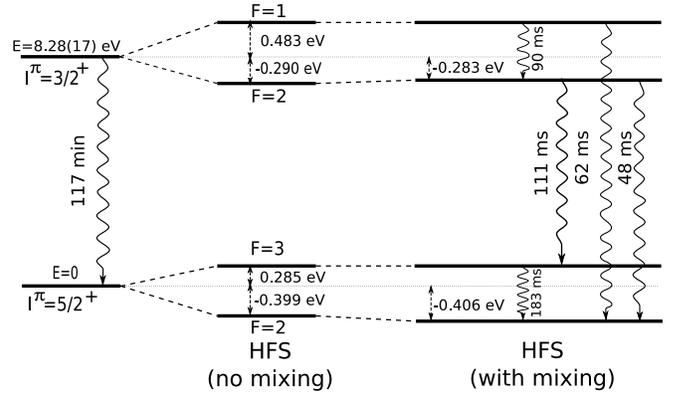}
\end{center}
\caption{Energy levels of $^{229}$Th$^{89+}$, g.s. and i.s.
  Left: the g.s. and i.s. of the bare thorium nucleus.
  Center: including ordinary hyperfine structure but neglecting NHM.
  Right: including hyperfine structure and NHM. For the displayed values $B(M1)=0.008$ W.u. was used.
The wavy lines accompanied by numbers indicate the radiative transitions
and the associated lifetimes.
  The scale is not maintained.}
\label{fig:thorium}
\end{figure}

Theoretical results for the hyperfine splitting (HFS) in H-, Li-, and B-like $^{229}$Th
are presented in Table~\ref{tab:totals} in terms of the matrix elements $V_{ik}$.
These values are obtained using the presently available experimental values
of the nuclear magnetic moments, $\mu^{(1)}=0.360(7)\mun$ for the g.s.
and $\mu^{(2)}=-0.37(6)\mun$ for the i.s. \cite{cam11,saf13,thi18},
where $\mun$ is the nuclear magneton.
The presented results have been calculated using in part
Refs.~\cite{sha94,sha97,sha99,bei00a,bei00b,sap06,ore08,gla10,gla19b}.
The details of the calculations are considered in the Supplemental Material.
Fig.~\ref{fig:thorium} shows the hyperfine structure of the g.s.
and i.s. of $^{229}$Th$^{89+}$ in absence (center) and including NHM (right).
As a reference, on the left side the levels for the bare nucleus are displayed.
For $B(M1)=0.008$ W.u. the NHM effect yields a matrix element of $V_{21}= -0.24$ eV
and shifts the ($3/2^+,F=2$) and ($5/2^+,F=2$) sublevels by 0.007 eV up and down, respectively. 
For the considered range of $B(M1)$ values from 0.005 to 0.048 W.u., the shift varies from 0.004 to 0.042 eV.
We note that the direct determination of $B(M1)$ from the g.s. HFS measurements is rather problematic
because of the large theoretical uncertainty originating from the nuclear magnetization
distribution correction (the Bohr-Weisskopf effect). The determination via the $g$ factor
considered below is much less sensitive to this effect.

The radiative $M1$ transition probability ${w_0}$ between i.s and g.s. in the bare $^{229}$Th nucleus is
($\hbar=c=1$\,,\;$\alpha = e^2/(4\pi)\,,\;e<0$)
\begin{equation}
\label{w0}
  w_0 = \frac{1}{4\pi}\,\frac{\omega^3}{3}\,d^2\,\mun^2
      = \alpha\,\frac{\omega^3}{12}\,\frac{d^2}{\massp^2}
\,,
\end{equation}
Here, $\omega$ is the transition frequency,
$\masse$ and $\massp$ are the electron and proton masses, respectively,
$d = \langle 3/2^+||\bmu^{(\text{n})}||5/2^+\rangle/\mun$
is the reduced matrix element of the nuclear magnetic moment operator $\bmu^{(\text{n})}$
between the i.s. and g.s., expressed in the nuclear magnetons. 
$d$ is directly related to the reduced transition probability $B(M1)$ in the Weisskopf units: $\BWu = d^2 / 30$.

Taking into account that the transition wavelength is much larger
than the size of the ion, the mixed $M1$ transition probability
between the $F$ and  $F^{\prime}$ states is given by
\begin{align}
  w_{F\rightarrow F'} &= \frac{1}{4\pi}\,\frac{4}{3}\,\omega^3\,\frac{1}{(2F+1)}
\nonumber\\
    & \sum_{M_{F},M_{F'}}|\langle \overline{F'M_{F'}jI'}|
      (\mbox{\boldmath$\mu$}^{(\text{e})}+\mbox{\boldmath$\mu$}^{(\text{n})})
      |\overline{FM_{F}jI}\rangle|^2
\,,
\end{align}
where $\mbox{\boldmath$\mu$}^{(\text{e})}$ is the magnetic moment operator
of the electronic subsystem and $\omega$ is the transition frequency.
In the case of the ($F=2$)$\rightarrow$ ($F^{\prime}=2$) transition,
the calculation using the Eckart-Wigner theorem yields
\begin{align}
\label{w1}
  &w_{(F=2,\text{up})\rightarrow (F^{\prime}=2,\text{low})} = \frac{1}{4\pi}\,\omega^3
\nonumber\\
    &\quad\times\frac{25}{18}\Bigg[b\sqrt{1-b^2}\left(\gfe\mub
      + \frac{14}{5}g_I^{(1)}\mun-\frac{9}{5}g_I^{(2)}\mun\right)
\nonumber\\
    &\qquad\qquad- \left(1-2b^2\right)\,\frac{\sqrt{2}}{5\sqrt{5}}\,d\,\mun\Bigg]^2
\,,
\end{align}
where $\gfe$ is the electronic $g$ factor, including the relativistic,
QED, interelectronic-interaction, and nuclear effects,
$g_I^{(1)}$ and $g_I^{(2)}$ are the nuclear $g$ factors ($\mu = g_I \mun I$)
of the g.s. and i.s., respectively, and $\mub$ is the Bohr magneton.
For the other transitions we find
\begin{align} \label{can98}
  &w_{(F=3,\text{low})\rightarrow (F^{\prime}=2,\text{low})} = \frac{1}{4\pi}\,\omega^3
\nonumber\\
  &\quad\times\frac{5}{9}
    \left[\sqrt{1-b^2}\left(\gfe\mub+g_I^{(1)}\mun\right)
      - b\,\frac{\sqrt{2}}{\sqrt{5}}\,d\,\mun\right]^2
\,,\\
  &w_{(F=1,\text{up})\rightarrow (F^{\prime}=2,\text{up})} = \frac{1}{4\pi}\,\omega^3
\nonumber\\
  &\quad\times\frac{5}{6}
    \left[\sqrt{1-b^2}\left(\gfe\mub+g_I^{(2)}\mun\right)
      + b\,\frac{\sqrt{2}}{\sqrt{5}}\,d\,\mun\right]^2
\,,\\
\label{w4}
  &w_{(F=2,\text{up})\rightarrow (F^{\prime}=3,\text{low})} = \frac{1}{4\pi}\,\omega^3
\nonumber\\
  &\quad\times\frac{7}{9}
    \left[b\left(\gfe\mub+g_I^{(1)}\mun\right)
      + \sqrt{1-b^2}\,\frac{\sqrt{2}}{\sqrt{5}}\,d\,\mun\right]^2
\,,\\
  &w_{(F=1,\text{up})\rightarrow (F^{\prime}=2,\text{low})} = \frac{1}{4\pi}\,\omega^3
\nonumber\\
  &\quad\times\frac{5}{6}
    \left[b\left(\gfe\mub+g_I^{(2)}\mun\right)
      - \sqrt{1-b^2}\,\frac{\sqrt{2}}{\sqrt{5}}\,d\,\mun\right]^2
\,.
\label{w5}
\end{align}
Except for Eq. (\ref{w4}), these equations are in agreement with those in Ref. \cite{kar98},
if we replace $\gfe$ with its one-electron Dirac value and neglect the contributions containing $\mun$.
As to the $(F=2,\text{up})\rightarrow (F^{\prime}=3,\text{low})$ transition, in Ref. \cite{kar98} the corresponding expression contains a coefficient of $7/6$ instead of $7/9$ as obtained here.
For $b=0$, Eq. (\ref{can98}) agrees with that from Ref. \cite{sha98}.

The M1 transition probabilities and the related lifetimes ($\tau=1/w$)
evaluated by formulas (\ref{w1})--(\ref{w5}) for $B(M1)$
in the range from 0.005 to 0.048 W.u. are listed in the Supplemental Material.
For $B(M1)=0.008$ W.u. the values of the lifetimes are presented in
Fig.~\ref{fig:thorium}. We note that the M1 transition probabilities between
the ``up'' and ``low'' states (Eqs. (\ref{w1}), (\ref{w4}), and (\ref{w5})) are approximately
linearly proportional to  $B(M1)$.

Let us consider an ion of $^{229}$Th with one valence electron exposed
to a homogeneous magnetic field $\bfB$ directed along the $z$ axis.
Assuming that the Zeeman splitting is much smaller than the hyperfine splitting,
$\DEmagn\ll \DEhfs$, the linear (in $B$) part of the energy shift can be written as
\begin{equation}\label{magsplit}
  \DEmagn = g \mub B \mf
\,,
\end{equation}
where $\mf$ is the $z$ projection of the total atomic angular momentum $F$.
We refer to the Supplemental Material for the details regarding the $g$-factor theory.
The $g$ factor of the ground $F=2$ state can be conveniently written as
\begin{equation}\label{gtot4}
  g = A b^2 + B d b\sqrt{1-b^2} + C\,,
\end{equation}
where the coefficients $A$, $B$, and $C$ do not depend on $b$ and $d$,
\begin{align}
\label{A}
  A &= \frac{5}{12}\,\gfe + \frac{\masse}{\massp}
       \left(\frac{7}{6}\,g_I^{(1)} - \frac{3}{4}\,g_I^{(2)}\right)
\,,\\
\label{B}
  B &= - \frac{1}{3\sqrt{10}}\,\frac{\masse}{\massp}
\,,\\
\label{C}
  C &= - \frac{1}{6}\,\gfe - \frac{7}{6}\,\frac{\masse}{\massp}\,g_I^{(1)} + \dghfs
\,.
\end{align}
The total theoretical values of the electronic $g$ factor, $\gfe$, 
for H-, Li-, and B-like thorium are presented in Table~\ref{tab:totals}.
They have been obtained using in part the results from
Refs.~\cite{bei00a,bei00b,sha02b,nef02,yer04,gla04,sha15,cak20,yer20}.
The last term in Eq. (\ref{C}) describes the HFS correction to the $g$ factor \cite{mos04,mos08,yer12,vol17}. 
Since this term is rather small, it can be evaluated at $b=0$.
The results of this evaluation (see the Supplemental Material) are presented in Table~\ref{tab:totals}.

The values of the coefficients $A$, $B$, and $C$ including their uncertainties,
which are mainly limited by the experimental input data, are given in Table~\ref{tab:totals}.
$B$ is presently known to a relative accuracy of $10^{-10}$.
The largest uncertainty of the coefficient $C$ is due to the second term in Eq.~(\ref{C}), 
it amounts to about $2\times 10^{-6}$ and stems from the g.s. nuclear magnetic moment.
The relative uncertainty of the coefficient $A$ does not exceed $7\times 10^{-5}$ 
and is mainly determined by the i.s. nuclear magnetic moment.

In Table~\ref{tab:eqb} the individual terms contributing to the $g$ factor (Eq.~(\ref{gtot4})) 
are given for several $B(M1)$ values in the range from 0.005 to 0.048 W.u.
Assuming a state-of-art experimental $g$-factor measurement,
we find the relative uncertainty $\delta b_\text{exp}$,
to which the NHM coefficient $b$ can be derived from the experiment.
The obtained value of $b$, together with the experimental value of the excitation energy $\DEnuc$ 
and the theoretical values of the g.s. and i.s. HFS, yields the matrix element $V_{21}$ using Eq. (\ref{b}).
Employing the relations between $d$, $V_{21}$, and $B(M1)$ (see the Supplemental Material) 
one can deduce $B(M1)$ with a few-percent accuracy,
which depends equally on the accuracy of $\DEnuc$ and the accuracy of the ratio $V_{21}/d$. 
The theoretical uncertainty of $V_{21}/d$ is mainly due to the Bohr-Weisskopf
correction to the HFS.

In the case of B-like thorium, the accuracy of $C$ is not high enough to determine $b$,
since the contribution of $Ab^2$ becomes comparable to the uncertainty of $C$.
However, let us consider the ratios of the HFS matrix elements $V_{ik}^{(2s)}$ of Li-like ions and
 $V_{ik}^{(2p)}$ of B-like ions to the ones  $V_{ik}^{(1s)}$ of H-like ions,
 \begin{equation}
\label{eta}
\eta_{ik}^{(2s/1s)} = V_{ik}^{(2s)}/V_{ik}^{(1s)}
\,,
\;\;\;
\eta_{ik}^{(2p/1s)} =  V_{ik}^{(2p)}/V_{ik}^{(1s)}
\,.
\end{equation}
These ratios can be calculated to a higher accuracy than the individual matrix elements
(see the Supplemental Material). This offers the opportunity in combined measurement 
for different charge states to determine improved values for $b$ and $\mu^{(1)}$.
The numerical values of $\eta_{ik}^{(2s/1s)}$ and $\eta_{ik}^{(2p/1s)}$ are given in Table~\ref{tab:totals},
the uncertainties are mainly due to the Bohr-Weisskopf effect.
Let us now rewrite Eq.~(\ref{gtot4}) in the form:
\begin{align}\label{gtotexp}
  \gfexp + \frac{1}{6}\,\gfeth &- \dghfs =
\nonumber \\
    &= A b^2 + B db\sqrt{1-b^2}
    - \frac{7}{6}\,\frac{\masse}{\massp}\,g_I^{(1)}
\,,
\end{align}
where $\gfexp$ is the experimental value of the total ground-state $g$ factor
and $\gfeth$ is the theoretical value of the electronic $g$ factor.
Considering the difference of these equations for different charge states,
we eliminate the last term on the right-hand side. 
The derived equation without $g_I^{(1)}$ can be used for a more precise determination 
of $b$ employing Eqs.~(\ref{eta}) and the following relations between $b$ and $V_{ik}$
\begin{eqnarray} \label{bb}
  b^2&=&b_0^2\frac{2}{1+4b_0^2+\sqrt{1+4b_0^2}}\approx b_0^2(1-3b_0^2)\,,\\
\label{bb0}
  b_0&=&\frac{V_{21}}{\DEnuc}
        \frac{1}{1+(V_{22}-V_{11})/\DEnuc}\,.
\end{eqnarray}
Using the obtained value of $b$ and Eq.~(\ref{gtotexp}) for one of the ions, 
we can determine $g_I^{(1)}$ (and hence $\mu^{(1)}$) to a higher accuracy.
Assuming the reasonable experimental uncertainty of $<10^{-7}$ for $g$-factor measurements
of both H-like $^{229}$Th$^{89+}$ and Li-like $^{229}$Th$^{87+}$, the accuracy of the coefficient $b$ 
can be improved by a factor of 10 compared to the one for $^{229}$Th$^{89+}$ only. 
Likewise, the experimental value for the magnetic moment $\mu^{(1)}$ can be improved tenfold 
compared to the currently accepted value. Similarly, one can use the $g$-factor experiments 
on Li- and B-like ions to refine the value of $\mu^{(1)}$.

Concluding, by a precise measurement of the $g$ factor in H- or Li-like ion of $^{229}$Th,
the much sought-after lifetime of the nuclear clock transition can be determined experimentally  
on a few percent level. Remarkably, to achieve this goal a measurement of the ion in its ground state 
can be used, meaning that the nuclear lifetime is determined completely without the nuclear decay.
The approach utilizes NHM which is very pronounced in the considered charge states.
The experimental accuracy of typical nowadays $g$-factor experiments is orders-of-magnitude higher 
than required by the proposed method. The complete formulas for the transition probabilities 
have been derived including relativistic, electron-electron correlation, QED, and nuclear contributions.
Further substantial improvements can be achieved if several charge states are compared. 
As a byproduct, the precise measurement of the ground-state nuclear magnetic moment is deduced.
In addition, in the course of such a measurement evidence for NHM in atomic ions can be obtained. 
NHM is a fascinating research topic by its own since it allows the manipulation of nuclear lifetimes 
by orders of magnitude simply by attachment or removal of a single electron. 
For example, in the He-like $^{229}$Th$^{88+}$ ion with paired electrons, the effect is absent, 
and the lifetime corresponds to the one of the bare nucleus, i.e., about 2 hours. 
Removal of one electron shortens the lifetime to a few tens of ms (H-like $^{229}$Th$^{89+}$ ion), 
while the attachment of an electron increases the lifetime to several seconds (Li-like $^{229}$Th$^{87+}$ ion).

%%%%%%%%%%%%%%%%%%%%%%%%%%%%%%%%%%%%%%%%%%%%%%%%%%%%%%%%%%%%%%%%%%%%

%%%%%%%%%%%%%%%%%%%%%%%%%%%%%%%%%%%%%%%%%%%%%%%%%%%%%%%%%%%%%%%%%%%%%%%%%%%%%%%%%
% \section{Acknowledgments}

This work was supported by RFBR and ROSATOM according to the research project No. 20-21-00098. 
The support by RFBR grants No. 20-02-00199, No. 19-32-90278, and No. 19-02-00974 is also acknowledged. 
C.B. acknowledges the support by the State of Hesse (cluster project ELEMENTS).

%
%
%%%%%%%%%%%%%%%%%%%%%%%%%%%%%%%%%%%%%%%%%%%%%%%%%%%%%%%%%%%%%%%%%%%%

\begin{widetext}

{
\setlength{\tabcolsep}{10pt}
\renewcommand{\arraystretch}{1.2}

\begin{table}
  \caption{The theoretical values of the hyperfine-interaction matrix elements $V_{ik}$,
    their ratios $\eta_{ik}$ defined by Eqs. (\ref{eta}),
    electronic $g$ factor $g_e$, the HFS correction to the $g$ factor,
    and the coefficients $A$, $B$, and $C$ for H-, Li-, and B-like $^{229}$Th ions. 
    The values of $\mu^{(1)}=0.360(7)\mun$ and $\mu^{(2)}=-0.37(6)\mun$
    \cite{cam11,saf13,thi18} are used.}
\label{tab:totals}
%\vspace{0.5cm}
\begin{tabular}{lr@{}lr@{}lr@{}lr@{}lr@{}l}
  \hline
  Contribution
&\multicolumn{2}{c}{$^{229}$Th$^{89+}$}
&\multicolumn{2}{c}{$^{229}$Th$^{87+}$}
&\multicolumn{2}{c}{$\eta_{ik}^{(2s/1s)}$}
&\multicolumn{2}{c}{$^{229}$Th$^{85+}$} 
&\multicolumn{2}{c}{$\eta_{ik}^{(2p/1s)}$}
\\
  \hline
$V_{11}/(\mu^{(1)}/\mun)$ [eV]
                   & $-$1.&109 (16)     & $-$0.&1833 (27)    &&   & $-$0.&06201 (31)   &&   \\
$V_{22}/(\mu^{(2)}/\mun) $ [eV]
                   & 0.&783 (14)        & 0.&1293 (25)       &&   & 0.&04412 (26)      &&   \\
$V_{11}$ [eV]
                   & $-$0.&399 (8)      & $-$0.&0660 (13)    & 0.&1653 (2)   & $-$0.&0223 (4)     & 0.&0559 (5)   \\
$V_{22}$ [eV]
                   & $-$0.&290 (47)     & $-$0.&0478 (77)    & 0.&1651 (2)   & $-$0.&0163 (26)    & 0.&0564 (6)   \\
$V_{21}/d$ [eV]
                   & $-$0.&498 (11)     & $-$0.&0823 (20)    & 0.&1652 (3)   & $-$0.&02796 (22)   & 0.&0562 (8)   \\
\hline
$\gfe$             &    1.&676 202 (3)  &    1.&920 397 (3)  &&   &    0.&585 842 (3)  &&   \\
\hline
$\dghfs$           &    0.&000 000 185 (11)  &    0.&000 000 053 6 (21)  &&   & $-$0.&000 005 11 (5)  &&   \\
\hline
$A$                &    0.&698 610 (16) &    0.&800 358 (16) &&   &    0.&244 293 (16) &&   \\
$B$                & $-$0.&000 057 41   & $-$0.&000 057 41   &&   & $-$0.&000 057 41   &&   \\
$C$                & $-$0.&279 458 (2)  & $-$0.&320 158 (2)  &&   & $-$0.&097 732 (2)  &&   \\
\hline
\end{tabular}
\end{table}

\setlength{\tabcolsep}{12pt}

\begin{table}
  \caption{The individual contributions to the right-hand side of Eq.~(\ref{gtot4}) for H-, Li-, and B-like ions of $^{229}$Th.
    The uncertainty of $C$ (Eq.~(\ref{C})) is defined by the ground-state nuclear magnetic moment
    while the uncertainty of $A$ (Eq.~(\ref{A})) is due to the isomeric-state nuclear magnetic moment.
    The NHM coefficient $b$ is evaluated for the given values of $B(M1)$ and $\DEnuc=8.28(17)$ eV \cite{sei19}
    using the approximate ($b_{0}$) and the exact ($b$) equations.
    Its uncertainty caused by the uncertainties of $\DEnuc$ and $V_{21}$ as well as the related uncertainties
    of the contributions $Ab^2$ and $Bdb\sqrt{1-b^2}$ are omitted.
    $\delta b_\text{exp}$ indicates the relative uncertainty of $b$, to which it can be determined from Eq.~(\ref{gtot4}),
    provided the experimental value of $g$ is measured to an accuracy higher than that of $C$.
  }
\label{tab:eqb}
%\vspace{0.5cm}
\begin{tabular}{r@{}lr@{}lr@{}lr@{}lr@{}lr@{}lr@{}l}
  \hline
 \multicolumn{2}{c}{$B(M1)$}
&\multicolumn{2}{c}{$b_{0}$}
&\multicolumn{2}{c}{$b$}
&\multicolumn{2}{c}{$Ab^2$}
&\multicolumn{2}{c}{$Bdb\sqrt{1-b^2}$}
&\multicolumn{2}{c}{$C$}
&\multicolumn{2}{c}{$\delta b_\text{exp}$}
\\
  \hline
&&\multicolumn{4}{c}{$^{229}$Th$^{89+}$} &&&&&&&&
\\
  \hline
    0.&005  & $-$0.&0230 & $-$0.&0230 & 0.&000368 & 0.&000001 & $-$0.&279458(2) & &$3\times 10^{-3}$  \\
    0.&008  & $-$0.&0291 & $-$0.&0290 & 0.&000589 & 0.&000001 & $-$0.&279458(2) & &$2\times 10^{-3}$  \\
    0.&015  & $-$0.&0398 & $-$0.&0397 & 0.&001102 & 0.&000002 & $-$0.&279458(2) & &$1\times 10^{-3}$  \\
    0.&030  & $-$0.&0563 & $-$0.&0560 & 0.&002193 & 0.&000003 & $-$0.&279458(2) & &$5\times 10^{-4}$  \\
    0.&048  & $-$0.&0712 & $-$0.&0707 & 0.&003490 & 0.&000005 & $-$0.&279458(2) & &$3\times 10^{-4}$  \\
%   0.&120  & $-$0.&1126 & $-$0.&1105 & 0.&008534 & 0.&000012 & $-$0.&279458(2) & &$1\times 10^{-4}$  \\
  \hline
&&\multicolumn{4}{c}{$^{229}$Th$^{87+}$} &&&&&&&&
\\
  \hline
    0.&005  & $-$0.&00384 & $-$0.&00384 & 0.&0000118 & 0.&0000001 & $-$0.&320158(2) & &$8\times 10^{-2}$  \\
    0.&008  & $-$0.&00486 & $-$0.&00486 & 0.&0000189 & 0.&0000001 & $-$0.&320158(2) & &$5\times 10^{-2}$  \\
    0.&015  & $-$0.&00665 & $-$0.&00665 & 0.&0000354 & 0.&0000003 & $-$0.&320158(2) & &$3\times 10^{-2}$  \\
    0.&030  & $-$0.&00941 & $-$0.&00941 & 0.&0000708 & 0.&0000005 & $-$0.&320158(2) & &$14\times 10^{-3}$  \\
    0.&048  & $-$0.&01190 & $-$0.&01190 & 0.&0001133 & 0.&0000008 & $-$0.&320158(2) & &$9\times 10^{-3}$  \\
%   0.&120  & $-$0.&01882 & $-$0.&01881 & 0.&0002831 & 0.&0000020 & $-$0.&320158(2) & &$4\times 10^{-3}$  \\
  \hline
&&\multicolumn{4}{c}{$^{229}$Th$^{85+}$} &&&&&&&&
\\
  \hline
    0.&005  & $-$0.&00131 & $-$0.&00131 & 0.&00000042 & 0.&00000003 & $-$0.&097732(2) & &  \\
    0.&008  & $-$0.&00165 & $-$0.&00165 & 0.&00000067 & 0.&00000005 & $-$0.&097732(2) & &  \\
    0.&015  & $-$0.&00226 & $-$0.&00226 & 0.&00000125 & 0.&00000009 & $-$0.&097732(2) & &  \\
    0.&030  & $-$0.&00320 & $-$0.&00320 & 0.&00000250 & 0.&00000017 & $-$0.&097732(2) & &  \\
    0.&048  & $-$0.&00405 & $-$0.&00405 & 0.&00000400 & 0.&00000028 & $-$0.&097732(2) & &  \\
%   0.&120  & $-$0.&00640 & $-$0.&00640 & 0.&00001000 & 0.&00000070 & $-$0.&097732(2) & &  \\
  \hline
\end{tabular}
\end{table}

}

\end{widetext}

%\narrowtext

%%%%%%%%%%%%%%%%%%%%%%%%%%%%%%%%%%%%%%%%%%%%%%%%%%%%%%%%%%%%%%%%%%%%%%%

%%%%%%%%%%%%%%%%%%%%%%%%%%%%%%%%%%%%%%%%%%%%%%%%%%%%%%%%%%%%%%%%%%%%%%%%

\end{document}

%%%%%%%%%%%%%%%%%%%%%%%%%%%%%%%%%%%%%%%%%%%%%%%%%%%%%%%%%%%%%%%%%%%%

\title{Supplemental Material:\\ 
  Ground-state $g$ factor of highly charged $^{229}$Th ions:\\
  an access to the M1 transition probability between the isomeric and ground nuclear states}

\author{V.~M.~Shabaev$^1$, D.~A.~Glazov$^1$, A.~M.~Ryzhkov$^1$, C.~Brandau$^{2,3}$, G.~Plunien$^4$,
W.~Quint$^3$, A.~M.~Volchkova$^1$, and D.~V.~Zinenko$^1$}

\affiliation{
$^1$Department of Physics, St. Petersburg State University, Universitetskaya 7/9, 199034 St. Petersburg, Russia\\
  $^2$I. Physikalisches Institut, Justus-Liebig-Universit\"at,  Heinrich-Buff-Ring 16,
  D-35392 Giessen, Germany\\
$^3$GSI Helmholtzzentrum f\"ur Schwerionenforschung GmbH,  Planckstrasse 1, D-64291 Darmstadt, Germany\\
$^4$Institut f\"ur Theoretische Physik, TU Dresden, Mommsenstrasse 13, Dresden, D-01062, Germany
}

\maketitle

In this Supplemental Material, we present the details concerning the theory
of the hyperfine and Zeeman splitting in H-, Li-, and B-like thorium ions.
The relativistic units ($\hbar=c=1$) and the Heaviside charge unit ($\alpha = e^2/(4\pi)\,,\;e<0$)
are used, $\mub=|e|/2\masse c$ is the Bohr magneton, $\mun=|e|/2\massp c$ is the nuclear magneton, 
$\masse$ and $\massp$ are the electron and proton masses, respectively.
 
The hyperfine splitting (HFS)  of a few-electron high-$Z$ ion with one valence electron  
can be written as
\begin{align} 
\label{eqhfs2}
  \Delta E_{F} =& \frac{\alpha(\alpha Z)^{3}}{n^{3}}\frac{\mu}{\mun}    
    \frac{\masse}{\massp}\frac{F(F+1)-I(I+1)-j(j+1)}{2Ij(j+1)(2l+1)}    
\nonumber\\    
    &\times \masse\,\Bigg\{\left[A(\alpha Z)(1-\delta)(1-\varepsilon) + \xrad\right]
\nonumber\\
    &+ \frac{B(\alpha Z)}{Z} + \frac{C(\alpha Z)}{Z^2} + \cdots \Bigg\}
\,,
\end{align}
where $Z$ is the nuclear charge number, $n$, $j$, and $l$ are the quantum numbers 
of the valence electron, $\mu$ is the nuclear magnetic moment.
The terms in the square brackets determine the one-electron contribution: 
$A(\alpha Z)$ is the one-electron relativistic factor, 
$\delta$ is the nuclear charge distribution correction, 
$\varepsilon$ is the nuclear magnetization distribution correction (so-called Bohr-Weisskopf effect),
and $\xrad$ is the QED correction. 
The last two terms in the braces refer only to ions with more than one electron. 
They describe the electron-electron interaction corrections of the first and second orders in $1/Z$, respectively.
The one-electron relativistic factor is given by \cite{pyy73,zap78}
\begin{equation}    
  A(\alpha Z) = \frac{n^{3}(2l+1)\kappa [2\kappa(\gamma+n_{r})-N]}{N^{4}\gamma(4\gamma^{2}-1)}\,,    
\end{equation}  
where $\kappa=(-1)^{j+l+1/2}(j+1/2)$ is the angular momentum-parity quantum number, 
$n_r=n-|\kappa|$ is the radial quantum number, 
$\gamma=\sqrt{\kappa^2-(\alpha Z)^2}$, and $N=\sqrt{n_r^2+2n_r\gamma+\kappa^2}$.
In the case of the nuclear hyperfine mixing (NHM), one should consider
the HFS matrix $V_{ik}$.
The diagonal HFS matrix elements $V_{11}$ and $V_{22}$ are just $\Delta E_{F}$ 
for the corresponding values of the quantum numbers,
\begin{align} 
\label{V11}
  V_{11} &= \Delta E_{F} [F=2,I^{\pi}=5/2^{+}]
\,,\\
\label{V22}
  V_{22} &= \Delta E_{F} [F=2,I^{\pi}=3/2^{+}]
\,.
\end{align}
The values of $A$, $\delta$, $\varepsilon$, $\xrad$, $B$, and $C$
for $^{229}$Th$^{89+}$, $^{229}$Th$^{87+}$, and $^{229}$Th$^{85+}$
are presented in Table \ref{tab:hfs}. The Bohr-Weisskopf effect is evaluated 
within the framework of the single-particle nuclear model \cite{sha94,sha97}.
The QED and interelectronic-interaction corrections are obtained using in part 
the results of Refs.~\cite{sha94,sha97,sha99,bei00a,bei00b,sap06,ore08,gla10,gla19b}.
Since the current precision of the nuclear magnetic moments is not high enough \cite{cam11,saf13,thi18}, 
in addition to the hyperfine shifts $V_{ii}$, we give the values of $V_{ii}/(\mu/\mun)$.
The values of $\Delta E_{F=3}$ and $\Delta E_{F=1}$, which are not affected by the NHM,
are also presented.

The nondiagonal matrix element of the HFS operator between the $F=2$ states
with $j=1/2$ can be written in the form:
\begin{align} \label{v21}
  V_{21} =& -\frac{\alpha(\alpha Z)^{3}}{n^{3}}\,\frac{\masse}{\massp}\,d
%    \frac{\langle 3/2^+||\vec{\mu}||5/2^+\rangle}{\mu_N}
    \,\frac{2\sqrt{2}}{3\sqrt{5}}\,\frac{1}{2l+1}     
\nonumber\\    
    &\times \masse \Bigg\{\left[A(\alpha Z)(1-\delta)(1-\varepsilon) + \xrad\right]
\nonumber\\
    &+ \frac{B(\alpha Z)}{Z} + \frac{C(\alpha Z)}{Z^2} + \cdots \Bigg\}
\,,
\end{align}
where $d = \langle 3/2^+||\bmu^{(\text{n})}||5/2^+\rangle/\mun$
is the reduced matrix element of the nuclear magnetic moment operator $\bmu^{(\text{n})}$ 
between the isomeric and ground nuclear states, expressed in the nuclear magnetons.
It is connected with the reduced transition probability $B(M1)$,
\begin{align} 
\label{bm1}
  B(M1) &\equiv B(M1;3/2^+\rightarrow 5/2^+) 
\nonumber\\
    &= \frac{3}{16\pi}|\langle 3/2^+||\bmu^{(\text{n})}||5/2^+\rangle|^2 
\,,
\end{align}
which is usually presented in the Weisskopf units, $B(M1) =(45/8\pi)\mun^2 \BWu$,
so that we have $\BWu=d^2/30$. 
Strictly speaking, the nuclear corrections in Eq.~(\ref{v21}) 
are not exactly the same as in Eq.~(\ref{eqhfs2}).
However, to a very good accuracy, we can neglect the difference 
in the nuclear charge distribution between the ground and isomeric nuclear states 
($\langle r^2\rangle^{(2)}-\langle r^2\rangle^{(1)}=0.012$ fm$^2$ \cite{thi18}) 
and, therefore, use the same values for this effect.
We also assume that the nuclear magnetization distribution correction in Eq.~(\ref{v21}) 
can be estimated by averaging the values of the related corrections to the HFS 
for the ground and isomeric states and summing quadratically their uncertainties.
The main problem consists in evaluation of the nuclear matrix element
$d$ (or $\BWu$), which requires using some microscopic nuclear models
\cite{dyk98,ruc06,tka15,min17,min19,min21}.
However, this value can be easily derived using the experimental value of
the nuclear excitation energy, $\DEnuc$,
if the mixing coefficient $b$ is known from experiment. 
In the present work, we show how $b$ can be determined from the $g$-factor experiments 
for the ground state of highly charged ions of $^{229}$Th.

Let us introduce the ratios
 of the HFS matrix elements $V_{ik}^{(2s)}$ of Li-like ions and
 $V_{ik}^{(2p)}$ of B-like ions to the ones  $V_{ik}^{(1s)}$ of H-like ions,
\begin{eqnarray}\label{eta}
\eta_{ik}^{(2s/1s)}=\frac{V_{ik}^{(2s)}}{V_{ik}^{(1s)}}\,,\;\;\;\;\;\eta_{ik}^{(2p/1s)}=\frac{V_{ik}^{(2p)}}{V_{ik}^{(1s)}}\,.
\end{eqnarray}  
It can be seen from Eqs. (\ref{eqhfs2}), (\ref{V11})-(\ref{v21}) that  the uncertainty due to the nuclear magnetic moments
and the nuclear transition parameter $d$  cancels out in these ratios.
Also, the uncertainty caused by the Bohr-Weisskopf (BW) effect is reduced in the ratio
$\eta_{ik}^{(2s/1s)}$ by an order of magnitude compared to the related uncertainty in   $V_{ik}^{(1s)}$
  and $V_{ik}^{(2s)}$. This is due to the fact that in the ratio we
  have
  \begin{eqnarray}
  (1-\veps^{(2s)})/(1-\veps^{(1s)})\approx 1- (\veps^{(2s)}-\veps^{(1s)})\,,
  \end{eqnarray}
      and the corrections $\veps^{(2s)}$ and $\veps^{(1s)}$ are strongly correlated
    with each other because of the same bahaviour  of the $1s$ and $2s$ electronic
    wave functions in the nuclear region (see, e.g., Ref. \cite{sha98}).
    This is also true for  $s$ and $p_{1/2}$ states at high values of the nuclear
    charge number $Z$ (see Ref. \cite{sha06} for details). However, in the case
    of the coefficient $\eta_{ik}^{(2p/1s)}$, the cancellation  of the BW effect is not
    so strong. Thus, the ratios $\eta_{ik}^{(2s/1s)}$ and $\eta_{ik}^{(2p/1s)}$
    can be calculated to a higher accuracy than that of the HFS matrix elements.
    As a result, the values $V_{ik}^{(2s)}$ and $V_{ik}^{(2p)}$ can be expressed
    with good accuracy in terms of $V_{ik}^{(1s)}$ and vice versa.

Next, we consider an ion of $^{229}$Th with one valence electron exposed
to a homogeneous magnetic field $\bfB$ directed along the $z$ axis.
Assuming that the Zeeman splitting is much smaller than the hyperfine splitting, 
$\DEmagn\ll \DEhfs$, the linear (in $B$) part of the energy shift can be written as
\begin{equation}\label{magsplit}
  \DEmagn = g \mub B \mf 
\,,
\end{equation}
where $\mf$ is the $z$ projection of the total atomic angular momentum $F$. 
To the lowest-order approximation, we have
\begin{equation} 
\label{de2}
  \DEmagn = \overline{\langle 5/2^+,F=2|} \left(\Vmagne + \Vmagnn\right) \overline{|5/2^+,F=2\rangle}
\,,
\end{equation}
where $\Vmagne$ is the interaction of the electrons with the homogeneous magnetic field,
\begin{equation}\label{W}
  \Vmagne = -e \sum_i(\balpha_i\cdot\bfA(\bfr_i))
          = \frac{|e|}{2}\,\left(\bfB \cdot \sum_i [\bfr_i \times \balpha_i]\right)
\,,
\end{equation}
$\balpha$ is the vector of the Dirac matrices, 
$\bfA(\bfr)=[\bfB\times\bfr]/2$ is the classical vector potential, 
and the summation runs over all electrons. The operator
\begin{equation}\label{Wn}
  \Vmagnn = - (\bmu \cdot \bfB)
\end{equation}
describes the interaction of the nuclear magnetic moment $\bmu$ with $\bfB$.
The relativistic and electron-electron correlation effects within the Breit approximation
can be obtained from Eq.~(\ref{de2}) using the proper many-electron wave functions. 
The QED, nuclear recoil and hyperfine-interaction corrections 
should be taken into account separately.
Substitution of the ground-state wave function defined by Eq.~(1)
of the main text of the Letter
into Eq.~(\ref{de2}) yields the following expression for the $g$ factor
\begin{align}
\label{gtot1}
  g =& - \frac{1}{6}\gfe\left(1-\frac{5}{2}b^2\right) 
       - \frac{7}{6}\frac{\masse}{\massp}g_I^{(1)}\left(1-b^2\right)
       - \frac{3}{4}\frac{\masse}{\massp}g_I^{(2)}b^2 
\nonumber\\
    &  - \frac{1}{3\sqrt{10}}\frac{\masse}{\massp}
        %  \frac{1}{\mu_N}\langle 3/2^+||\vec{\mu}||5/2^+\rangle b\sqrt{1-b^2}
         d b\sqrt{1-b^2}
% \nonumber\\
%     &  
    + \dghfs
\,,
\end{align}
where the difference between the electronic $\gfe$ factors for the ground 
and isomeric nuclear states, which is mainly due to the nuclear size effect, is neglected.
The first term of Eq.~(\ref{gtot1}) comes from the expectation value of the electronic 
operator $\Vmagne$ (\ref{W}) with the first and second components of the wave function
(see Eq.~(1) of the main text of the Letter).
It includes also all the corrections to the electronic $\gfe$ factor, 
which are beyond the approximation defined by Eq.~(\ref{de2}).
The second and third terms result from the nuclear operator $\Vmagnn$ (\ref{Wn}).
The nuclear $g$ factors of the ground ($g_I^{(1)}$) and isomeric ($g_I^{(2)}$)
states are defined by $\mu = g_I \mun I$. 
The fourth term of Eq.~(\ref{gtot1}) is due to the interference of the first and second
components of the wave function which takes place for the operator $\Vmagnn$. 
Finally, the last term describes the HFS correction 
to the $g$ factor~\cite{mos04,mos08,yerokhin:12:pra,volchkova:17:nimb}. 
Assuming that $b$ is rather small, in what follows, we consider the HFS correction for $b=0$.

Table~\ref{tab:g} presents the individual contributions to the electronic $g$ factors 
of the ground state of H-, Li-, and B-like $^{229}$Th ions.
We use the results presented in Refs.~\cite{bei00a,bei00b,sha02b,yer04,gla04,cak20,yer20} 
(see also references therein). In most of these papers, there are no values for $Z=90$.
So, we evaluate the leading contributions and interpolate the data for the higher-order corrections.
To estimate the nuclear polarization effect we use the corresponding values for $^{232}$Th ions
evaluated in Ref.~\cite{nef02} and assume an uncertainty of 100\%.
The nuclear polarization contribution defines generally
the ultimate limit for the theoretical accuracy.
However, at present the total theoretical uncertainties are mainly determined by the unknown 
higher-order terms of the two-loop QED, screened QED, and interelectronic-interaction contributions.

The hyperfine-interaction correction $\dghfs$ is conveniently written in the form
\begin{align}
  \dghfs &= \dghfsm + \dghfsQ
\,,\\
  \dghfsm &= k_{\mu}(\masse/\massp) g_I^{(1)}
\,,\\
  \dghfsQ &= k_Q (\masse c/\hbar)^2 Q
\,,
\end{align}  
where $Q$ is the nuclear electric quadrupole moment.
We derive the coefficients $k_{\mu}$ and $k_Q$ from the calculations 
of Refs.~\cite{mos04,mos08} for H- and Li-like thorium.
For $^{229}$Th$^{85+}$ we use the data from Ref.~\cite{volchkova:17:nimb}
for $k_{\mu}$ and calculate $k_Q$ here using the same methods.
The QED correction to $\dghfsm$ was calculated for $1s$ state only~\cite{yerokhin:12:pra}.
Since for $\dghfsQ$ it is unknown, we do not take it into account here.
The results are presented in Table~\ref{tab:g_hfs}, where $\dghfsm$
and $\dghfsQ$ are obtained using the following values 
of the nuclear dipole magnetic and quadrupole electric moments of $^{229}$Th:
$\mu/\mun=0.360(7)$ \cite{cam11,saf13} and $Q=3.15(3)\times 10^{-28}$m$^2$
\cite{bem88} (see also Ref. \cite{saf13}).

Finally, in Table~\ref{tab:trans}, the M1 transition probabilities ($w$) between the HFS levels 
and the related lifetimes ($\tau=1/w$), for $B(M1)$ in the range from 0.005 to 0.048 W.u. are listed.
These values are obtained by formulas (9)--(13) from the main text of the Letter.
The corresponding transition energies depend on the HFS shifts $\Delta E_{F}$ (see Table~\ref{tab:hfs})
and the nuclear isomeric state energy, which was taken from Ref.~\cite{sei19},  $\DEnuc=8.28(17)$~eV.
For the energies counted from the center of gravity of the $I=5/2^{+}$ ordinary (no mixing) HFS multiplet,
we have $E_{F=3}=\Delta E_{F=3}$, $E_{F=1}=\Delta E_{F=1}+\DEnuc$.
The energies of the $F=2$ states, $E_{F=2,\text{low}}$ and $E_{F=2,\text{up}}$, which include the mixing effect,
are given by the formula~(4) from the main text of this Letter, where $E_1^0=0$ and $E_2^0=\DEnuc$.
Their numerical values are presented in Table~\ref{tab:trans}.

\begin{figure}
\begin{center}
  \includegraphics[width=\columnwidth]{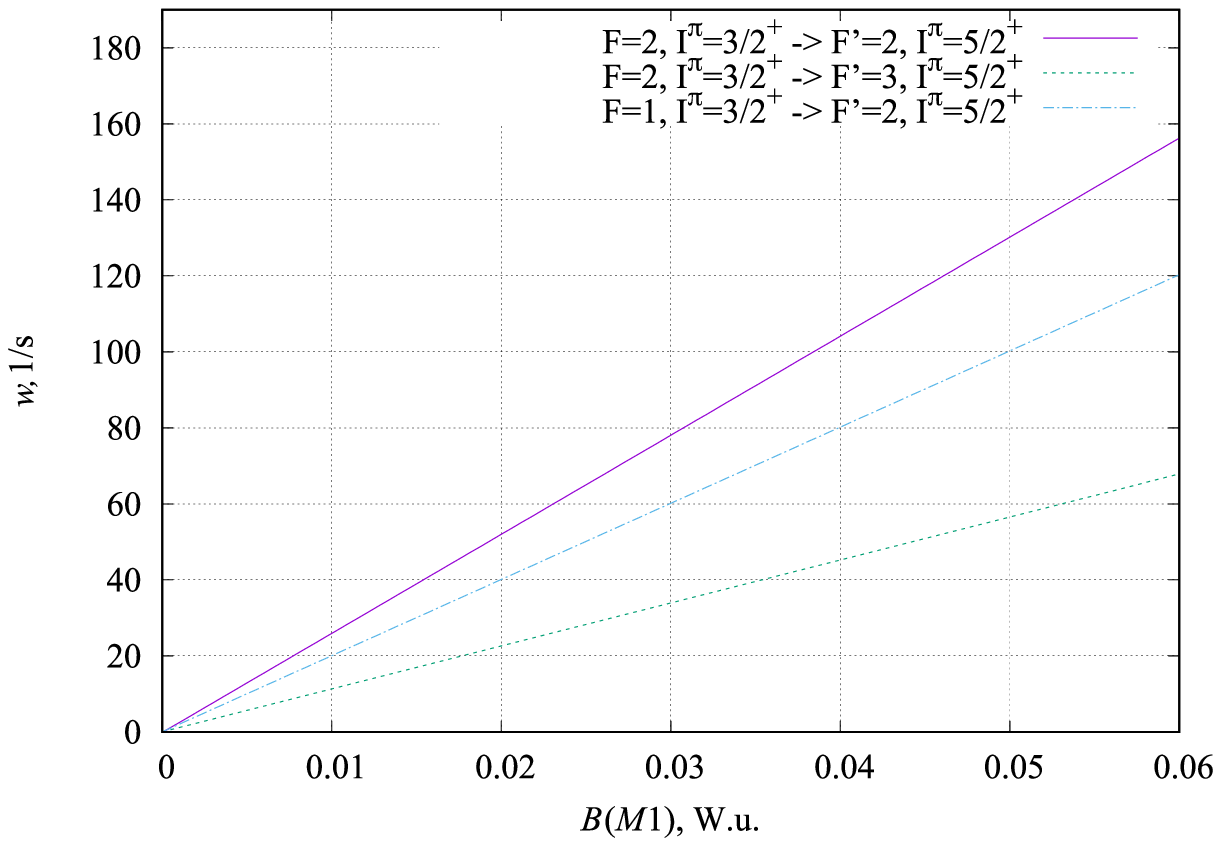}
\end{center}
\caption{The probabilities of the ($F$,~$I^{\pi}=3/2^{+}$)$\rightarrow$($F'$,~$I^{\pi}=5/2^{+}$)
%  ($F$, $I=3/2$)$\rightarrow$  ($F'$, $I=5/2$)
transitions  as functions of  $B(M1)$
in  $^{229}$Th$^{89+}$.}
\label{fig:thorium_sm}
\end{figure}

In Fig.~\ref{fig:thorium_sm} we present the probabilities of
the ($F$,~$I^{\pi}=3/2^{+}$)$\rightarrow$($F'$,~$I^{\pi}=5/2^{+}$)
transitions as functions of  $B(M1)$
for $^{229}$Th$^{89+}$.
This figure clearly demostrates the nearly linear proportionality
of the probability to $B(M1)$ for the transitions
between the up ($I^{\pi}=3/2^{+}$) and low  ($I^{\pi}=5/2^{+}$)  states. The same
is true for the corresponding transitions in  $^{229}$Th$^{87+}$ and $^{229}$Th$^{85+}$.
The linear dependence means that for all $B(M1)$
under consideration the transition chanel due to
the NHM remains dominant.

%%%%%%%%%%%%%%%%%%%%%%%%%%%%%%%%%%%%%%%%%%%%%%%%%%%%%%%%%%%%%%%%%%%%

\begin{widetext}

{
\setlength{\tabcolsep}{12pt}
\renewcommand{\arraystretch}{1.2}

\begin{table}
  \caption{The individual contributions to the ground-state HFS
    of H-, Li-, and B-like $^{229}$Th ions, neglecting the mixing effect.
    The values of $\mu^{(1)}/\mun=0.360(7)$ and $\mu^{(2)}/\mun=-0.37(6)$ 
    \cite{cam11,saf13,thi18} are used.}
\label{tab:hfs}
%\vspace{0.5cm}
\begin{tabular}{lr@{}lr@{}lr@{}l}
  \hline
  Contribution 
&\multicolumn{2}{c}{$^{229}$Th$^{89+}$} 
&\multicolumn{2}{c}{$^{229}$Th$^{87+}$}  
&\multicolumn{2}{c}{$^{229}$Th$^{85+}$} \\
  \hline  
$A(\alpha Z)$      &    2.&60940     &    3.&63900     &    3.&31732      \\
$\delta$           &    0.&168 (1)   &    0.&181 (1)   &    0.&064        \\
$\eps^{(1)}$       &    0.&042 (14)  &    0.&046 (15)  &    0.&014 (5)    \\
$\eps^{(2)}$       &    0.&053 (18)  &    0.&058 (19)  &    0.&018 (6)    \\
$\xrad$            & $-$0.&014 (1)   & $-$0.&019 (1)   & $-$0.&003        \\
$ [B(\alpha Z)/Z + C(\alpha Z)/Z^2]^{(1)}$   
                   &      &          & $-$0.&0943      & $-$0.&288        \\
$ [B(\alpha Z)/Z + C(\alpha Z)/Z^2]^{(2)}$   
                   &      &          & $-$0.&0931      & $-$0.&286        \\  
\hline
$V_{11}/(\mu^{(1)}/\mun)$ [eV]  
                   & $-$1.&109 (16)  & $-$0.&1833 (27) & $-$0.&06201 (31) \\
$V_{22}/(\mu^{(2)}/\mun)$ [eV]  
                   &    0.&783 (14)  &    0.&1293 (25) &    0.&04412 (26) \\
$V_{11}$ [eV]  
                   & $-$0.&399 (10)  & $-$0.&0660 (16) & $-$0.&0223 (4)   \\
$V_{22}$ [eV] 
                   & $-$0.&290 (47)  & $-$0.&0478 (77) & $-$0.&0163 (26)  \\  
$V_{21}/d$ [eV]  
                   & $-$0.&498 (11)  & $-$0.&0823 (20) & $-$0.&02796 (22) \\
\hline
$\Delta E_{F=3}$ [eV]  
                   &    0.&285 (7)   &    0.&047 (1)   &    0.&016        \\
$\Delta E_{F=1}$ [eV]  
                   &    0.&483 (79)  &    0.&080 (13)  &    0.&027 (4)    \\
\hline
\end{tabular}
\end{table}

\begin{table}
  \caption{The individual contributions to the electronic $\gfe$ factor 
    of the ground state of H-, Li-, and B-like $^{229}$Th ions.}
\label{tab:g}
\begin{center}
\begin{tabular}{lr@{}lr@{}lr@{}l}
  \hline
  Contribution 
&\multicolumn{2}{c}{$^{229}$Th$^{89+}$} 
&\multicolumn{2}{c}{$^{229}$Th$^{87+}$}  
&\multicolumn{2}{c}{$^{229}$Th$^{85+}$} \\
  \hline
Dirac value (point nucleus) &    1.&672 130 874     &    1.&915 345 313     &    0.&582 011 979  \\
Finite nuclear size         &    0.&001 028 6       &    0.&000 191 5       &    0.&000 021 2    \\
One-loop QED                &    0.&003 045 0       &    0.&002 439 5       & $-$0.&000 385 9    \\
Two-loop QED                & $-$0.&000 003 8       & $-$0.&000 003 6       &    0.&000 001 2    \\
Interelectronic interaction $\sim 1/Z$      
                            &      &                &    0.&002 434 0       &    0.&004 235      \\
Interelectronic interaction $\sim 1/Z^{2+}$ 
                            &      &                & $-$0.&000 008 3       & $-$0.&000 022      \\
Screened QED                &      &                & $-$0.&000 002 1       & $-$0.&000 018      \\
Nuclear recoil              &    0.&000 002 3       &    0.&000 000 5       & $-$0.&000 001 5    \\
Nuclear polarization        & $-$0.&000 001 0       & $-$0.&000 000 2       & $-$0.&000 000 02   \\
Total                       &    1.&676 202 (3)     &    1.&920 397 (3)     &    0.&585 842 (3)  \\
\hline
\end{tabular}
\end{center}
\end{table}

\begin{table}
  \caption{The hyperfine-interaction correction $\dghfs$ 
    to the ground-state $g$ factor of H-, Li-, and B-like $^{229}$Th ions.
    The following values of the nuclear moments are used: 
    $\mu/\mun=0.360(7)$ \cite{cam11,saf13} and $Q=3.15(3)\times 10^{-28}$m$^2$ 
    \cite{bem88,saf13}.}
\label{tab:g_hfs}
\begin{center}
\begin{tabular}{lr@{}lr@{}lr@{}l}
  \hline
  Contribution 
&\multicolumn{2}{c}{$^{229}$Th$^{89+}$} 
&\multicolumn{2}{c}{$^{229}$Th$^{87+}$}  
&\multicolumn{2}{c}{$^{229}$Th$^{85+}$} \\
  \hline
$k_{\mu}$       &    0.&00632       &    0.&00125       &    0.&00262   \\
$\dghfsm\times 10^6$        
                &    0.&496 (10)    &    0.&0980 (20)   &    0.&205 (4)  \\
$k_Q$           & $-$0.&000147      & $-$0.&0000210     & $-$0.&00251   \\
$\dghfsQ\times 10^6$        
                & $-$0.&311 (3)     & $-$0.&0444 (4)    & $-$5.&31 (5)  \\
$\dghfs\times 10^6$        
                &    0.&185 (11)    &    0.&0536 (21)   & $-$5.&11 (5)  \\
\hline
\end{tabular}
\end{center}
\end{table}

% \begin{widetext}
\setlength{\tabcolsep}{4pt}

\begin{table}
  \caption{The energies (in eV) of the HFS levels with $F=2$ counted 
    from the center of gravity of the $I=5/2^+$ ordinary HFS multiplet, 
    the $M1$ transition probabilities (in s$^{-1}$) and the corresponding lifetimes 
    ($\tau=1/w$, in s) calculated by Eqs.~(9)--(13) from the main text of this Letter 
    for the given values of $B(M1)$ (in W.u.) and $\DEnuc=8.28(17)$ eV \cite{sei19}. 
    The energies of the $F=1$ and $F=3$ states, which do not depend on the mixing effect, 
    can be found from $\Delta E_{F=1,3}$ (see Table~\ref{tab:hfs}) and $\DEnuc$ (see the text).
  }
\label{tab:trans}
%\vspace{0.5cm}
\begin{tabular}{r@{}lr@{}lr@{}lr@{}lr@{}lr@{}lr@{}lr@{}lr@{}lr@{}lr@{}lr@{}lr@{}l}
  \hline
 &&&&&
&\multicolumn{4}{c}{${2\text{up}\rightarrow 2\text{low}} $}
&\multicolumn{4}{c}{${3\text{low}\rightarrow 2\text{low}} $}
&\multicolumn{4}{c}{${1\text{up}\rightarrow 2\text{up}} $}
&\multicolumn{4}{c}{${2\text{up}\rightarrow 3\text{low}} $}
&\multicolumn{4}{c}{${1\text{up}\rightarrow 2\text{low}} $}
\\
 \multicolumn{2}{c}{$B(M1)$} 
&\multicolumn{2}{c}{$E_{F=2,\text{low}}$}
&\multicolumn{2}{c}{$E_{F=2,\text{up}}$}
&\multicolumn{2}{c}{$w$}
&\multicolumn{2}{c}{$\tau$}
&\multicolumn{2}{c}{$w$}
&\multicolumn{2}{c}{$\tau$}
&\multicolumn{2}{c}{$w$}
&\multicolumn{2}{c}{$\tau$}
&\multicolumn{2}{c}{$w$}
&\multicolumn{2}{c}{$\tau$}
&\multicolumn{2}{c}{$w$}
&\multicolumn{2}{c}{$\tau$}
\\
\hline
&&\multicolumn{4}{c}{$^{229}$Th$^{89+}$} &&&&&&&&&&&&&&&&&&&
\\
    0.&005  & $-$0.&403  & $\DEnuc-$0.&286  &  &12.9  & 0.&077  &   5.&40  & 0.&185 &  1&1.3  & 0.&089 &   &5.64 & 0.&177  &  &10.0  & 0.&100  \\
    0.&008  & $-$0.&406  & $\DEnuc-$0.&283  &  &20.7  & 0.&0483 &   5.&46  & 0.&183 &  1&1.2  & 0.&090 &   &9.04 & 0.&111  &  &16.0  & 0.&0623 \\ 
    0.&015  & $-$0.&412  & $\DEnuc-$0.&277  &  &38.9  & 0.&0257 &   5.&61  & 0.&178 &  1&0.9  & 0.&092 &  1&7.0  & 0.&0590 &  &30.1  & 0.&0332 \\
    0.&030  & $-$0.&426  & $\DEnuc-$0.&263  &  &78.0  & 0.&0128 &   5.&92  & 0.&169 &  1&0.3  & 0.&097 &  3&3.9  & 0.&0295 &  &60.2  & 0.&0166 \\
    0.&048  & $-$0.&441  & $\DEnuc-$0.&248  & 1&25.   & 0.&0080 &   6.&32  & 0.&158 &   &9.66 & 0.&104 &  5&4.3  & 0.&0184 &  &96.2  & 0.&0104 \\
\hline
&&\multicolumn{4}{c}{$^{229}$Th$^{87+}$} &&&&&&&&&&&&&&&&&&& 
\\
    0.&005  & $-$0.&066  & $\DEnuc-$0.&048  &  0.&455 &  2.&20  &  0.&0316 & &31.6  &  0.&067 & &14.8  &  0.&255 &  3.&92  &  0.&278 &  3.&60  \\
    0.&008  & $-$0.&066  & $\DEnuc-$0.&048  &  0.&728 &  1.&37  &  0.&0317 & &31.6  &  0.&067 & &14.9  &  0.&409 &  2.&45  &  0.&444 &  2.&25  \\ 
    0.&015  & $-$0.&066  & $\DEnuc-$0.&047  &  1.&37  &  0.&732 &  0.&0318 & &31.4  &  0.&067 & &14.9  &  0.&766 &  1.&31  &  0.&832 &  1.&20  \\
    0.&030  & $-$0.&067  & $\DEnuc-$0.&047  &  2.&73  &  0.&366 &  0.&0321 & &31.1  &  0.&066 & &15.1  &  1.&53  &  0.&653 &  1.&67  &  0.&601 \\
    0.&048  & $-$0.&067  & $\DEnuc-$0.&047  &  4.&37  &  0.&229 &  0.&0325 & &30.8  &  0.&066 & &15.2  &  2.&45  &  0.&408 &  2.&66  &  0.&375 \\
\hline
&&\multicolumn{4}{c}{$^{229}$Th$^{85+}$} &&&&&&&&&&&&&&&&&&&
\\
    0.&005  & $-$0.&022  & $\DEnuc-$0.&016  & 0.&0046 & 2&18.3  & 0.&000113 & &8832  & 0.&000249 & &4017  & 0.&00374 & 2&67.4  & 0.&00204 & 4&90.5  \\
    0.&008  & $-$0.&022  & $\DEnuc-$0.&016  & 0.&0073 & 1&36.5  & 0.&000113 & &8826  & 0.&000249 & &4019  & 0.&00598 & 1&67.1  & 0.&00326 & 3&06.6  \\ 
    0.&015  & $-$0.&022  & $\DEnuc-$0.&016  & 0.&0137 &  &72.77 & 0.&000113 & &8812  & 0.&000248 & &4025  & 0.&0112  &  &89.12 & 0.&00612 & 1&63.5  \\
    0.&030  & $-$0.&022  & $\DEnuc-$0.&016  & 0.&0275 &  &36.39 & 0.&000114 & &8783  & 0.&000248 & &4036  & 0.&0224  &  &44.56 & 0.&0122  &  &81.76 \\
    0.&048  & $-$0.&022  & $\DEnuc-$0.&016  & 0.&0440 &  &22.74 & 0.&000114 & &8748  & 0.&000247 & &4051  & 0.&0359  &  &27.85 & 0.&0196  &  &51.10 \\
%   \\
  \hline
\end{tabular}
\end{table}

}

\end{widetext}

%\narrowtext

%%%%%%%%%%%%%%%%%%%%%%%%%%%%%%%%%%%%%%%%%%%%%%%%%%%%%%%%%%%%%%%%%%%%%%%

%%%%%%%%%%%%%%%%%%%%%%%%%%%%%%%%%%%%%%%%%%%%%%%%%%%%%%%%%%%%%%%%%%%%%%%%